# Staggered field driven domain walls motion in antiferromagnetic heterojunctions

Y. L. Zhang, Z. Y. Chen, Z. R. Yan, D. Y. Chen, Z. Fan, and M. H. Qin [*]

*Institute for Advanced Materials and Guangdong Provincial Key Laboratory*
*of Quantum Engineering and Quantum Materials, South China Normal University,*
*Guangzhou 510006, China*

**[Abstract]** In this work, we study the antiferromagnetic (AFM) spin dynamics in heterostructures which consist of two kinds of AFM layers. Our micromagnetic simulations demonstrate that the AFM domain-wall (DW) can be driven by the other one (driven by field-like Néel spin-orbit torque, Phys. Rev. Lett. 117, 017202 (2016)) through the interface couplings. Furthermore, the two DWs detach from each other when the torque increases above a critical value. The critical field and the highest possible velocity of the DW depending on several factors are revealed and discussed. Bases on the calculated results, we propose a method to modulate efficiently the multi DWs in antiferromagnet, which definitely provides useful information for future AFM spintronics device design.

---

[*] Electronic mail: qinmh@scnu.edu.cn

Antiferromagnets are attracting more and more attentions due to their potential applications in the field of antiferromagnetic (AFM) spintronics. On the one hand, replacing ferromagnets by antiferromagnets in spintronics devices offers insensitivity to external magnetic field perturbations and produces no perturbing stray fields due to the zero net magnetic moment in the AFM element.[1-3] Thus, information stored in AFM domains or domain walls (DWs) is robust, and the AFM elements can be arranged with high density.[4] On the other hand, antiferromagnets also exhibit ultrafast spin dynamics with characteristic frequencies in the THz range.[5] More importantly, the velocity of AFM domain wall (DW) is only limited by the group velocity of spin waves,[6-8] which is almost two orders of magnitude larger than velocity of typical ferromagnetic (FM) DW which is limited by the Walker breakdown.[9-12] As a result, AFM spintronics is very promising in future storage devices, and several efficient methods of modulating AFM domains and driving DWs motion have been revealed.[13-22]

Recent experiments revealed that an applied electrical current can induce the local staggered effective field (or field-like Néel spin-orbit torque, NSOT)[23,24] in CuMnAs[25-27] and $Mn_2Au$[28] due to the spin-orbit effect and in turn modulates the orientation of the AFM moments. More interestingly, a high velocity ~ 10-100 km/s of the AFM DW motion driven by the NSOT has been predicted in theory.[13] Specifically, for an AFM DW under the staggered field $B_{1N}/B_{2N}$ coupling to the spin $\mathbf{m}_1/\mathbf{m}_2$, the torque $\Gamma_{1N}/\Gamma_{2N}$ is induced and cants $\mathbf{m}_1/\mathbf{m}_2$ forward, as depicted in Fig. 1(a). Subsequently, a rather large precession torque from the strong exchange interaction $\Gamma_{ex}^p$ is generated due to the spin deviation, and drives the DW motion. Furthermore, AFM DW motion has also been theoretically predicted by several external stimuli including spin wave excitations,[7,21] spin-orbit torques,[6,19] temperature gradient,[15-17] and asymmetric field pulses.[14,20] For example, it has been proven that the competition between the entropic torque and the Brownian force determines the AFM DW motion towards the hotter or colder region under an applied temperature gradient.[15-17]

It is noted that one of the major challenges in future AFM spintronics applications is fast transport of multi AFM DWs for information storage. Thus, these proposed methods do provide important information for applications, while several shortcomings deserve to be overcome. For example, NSOT only arises in these AFM materials with particular crystal structures, which is not available in most of antiferromagnets. More importantly, NSOT drives the neighboring AFM DWs approach to each other and annihilates them, which strongly hiders its application in future device where efficient motions of multi DWs are indispensable. For the other proposed driven methods, the drift velocities are not so large, and strict restrictions on the external stimuli are suggested. For example, our earlier work has demonstrated that considerable velocity of the AFM DW only can be obtained near the resonance frequency of the oscillation magnetic field in cooperation with a static field.[14] In addition, it has been proven in earlier work that the interplay between AFM and FM DWs in the FM-AFM double layers can shift the AFM DW, when the FM DW is driven by a spin current.[18] However, the velocity is also limited by the Walker breakdown. Thus, there is still an urgent need in searching for efficient methods of driving multi AFM DWs motion with a high speed.

Based on these earlier studies, we investigate the AFM dynamics in exchange coupled AFM1-AFM2 heterojunction in which the DW in AFM1 layers is driven by the NSOT, as simplistically depicted in Fig. 1(b). We figure out that the AFM2 DW can be driven efficiently by the AFM1 DW through the interface coupling. Furthermore, the highest possible velocity of the AFM2 DW relevant to several factors has been clarified and detailed explained. Based on this property, we put forward a proposal for controlling multi-DWs in antiferromagnet.

In this work, the heterojunction system is considered to be a three-dimensional cuboidal lattice with the free boundary conditions applied in the *x*-, *y*- and *z*-axis directions. The model Hamiltonian can be given by

$$H = H_{AF1} + H_{AF2} + H_{inter},  \qquad (1)$$

where

$$H_{AF1} = J_{AF1} \sum_{\langle i,j \rangle} \mathbf{m}_i \cdot \mathbf{m}_j - d_x \sum_i \left(\mathbf{m}_i^x\right)^2 - d_z \sum_i \left(\mathbf{m}_i^z\right)^2 + B_N \sum_i \mathbf{m}_i^x, \qquad (2)$$

$$H_{AF2} = J_{AF2} \sum_{\langle n,m \rangle} \mathbf{m}_n \cdot \mathbf{m}_m - D_x \sum_n \left(\mathbf{m}_n^x\right)^2 - D_z \sum_n \left(\mathbf{m}_n^z\right)^2, \qquad (3)$$

$$H_{\text{inter}} = J_{\text{inter}} \sum_{\langle i,n \rangle} \mathbf{m}_i \cdot \mathbf{m}_n . \tag{4}$$

where $J_{AF1}/J_{AF2}$ is the antiferromagnetic exchange interaction between the nearest neighbors in AFM1/AFM2 layers, and $J_{\text{inter}}$ is the interface coupling between the nearest neighbors, $\mathbf{m}_k = \mu_k/\mu_s$ ($k = i, j, m, n$) is the normalized magnetic moment at site $k$ with the saturated moment $\mu_s$, $m_k^r$ ($r = x, y, z$) is the $r$ component of $\mathbf{m}_k$, $d_x/J > d_z/J$ and $D_z/J > D_x/J$ are anisotropy constants, $B_N = B_{1N}$ ($B_{2N}$) is the staggered field along the $x$ ($-x$)-axis direction on the magnetic sublattice 1 (2) in AFM1 layers. Here, the magnitudes of the anisotropies could be modulated through tuning the thickness of the heterojunction. The simulation is performed on a $L_x \times L_y \times L_z$ (20 × 30 × 350, including 10 AFM1 layers and 10 AFM2 layers) lattice by solving the Landau-Lifshitz-Gilbert equation,[29,30]

$$\frac{\partial \mathbf{m}_i}{\partial t} = -\frac{\gamma}{(1+\alpha^2)} \mathbf{m}_i \times \left[ \mathbf{H}_i + \alpha \left( \mathbf{m}_i \times \mathbf{H}_i \right) \right], \tag{5}$$

where $\gamma$ is the gyromagnetic ratio, $\alpha = \alpha_1$ ($\alpha_2$) is the Gilbert damping constant in AFM1 (AFM2) layers, and the internal field is $\mathbf{H}_i = -\partial H/\partial \mathbf{m}_i$. Unless stated elsewhere, $J_{AF1} = J_{AF2} = J$ ($J = 1$ is the energy unit), $J_{\text{inter}} = 0.5J$, $d_x = D_z = 0.1J$, and $\alpha_1 = \alpha_2 = 0.01$ are chosen. Moreover, the fourth-order Runge-Kutta method is used to solve the equation with a time step $\Delta t = 5.0 \times 10^{-4} |\mu_s/\gamma J|$. After sufficient relaxation of the Néel-type DWs, we apply the staggered field and study the DWs motion. The local staggered magnetization $2\mathbf{n} = \mathbf{m}_1 - \mathbf{m}_2$ is calculated to describe the spin dynamics.

First, we study the case of the uniaxial anisotropy ($d_z = D_x = 0$). The initial spin configuration is presented in Fig. 2(a) which clearly shows two Néel-type AFM DWs. Specifically, AFM1 DW at the position $z = 30a$ and AFM2 DW at the position $z = 60a$ ($a$ is the lattice constant) are observed, as clearly shown in Fig. 2(b) which gives the three components of the local magnetization $\mathbf{n}_1$ and $\mathbf{n}_2$. Furthermore, the $y$ component of the magnetization in the whole system equals to zero $n_1^y = n_2^y = 0$, and a small $n_1^z$ ($n_2^x$) of the local magnetization in AFM1 (AFM2) domains is observed due to the interface coupling.

The induced $B_N$ by applied electrical current can drive the motion of the AFM1 DW. The velocity of the DW increases with the increase of $B_N$, as clearly shown in Fig. 2(c) which gives the displacements of the AFM1 DW (solid lines) and AFM2 DW (dashed lines) as functions of time for various $B_N$. Similar to the earlier report, the AFM1 DW can also shift the AFM2 DW at the expense of its velocity through the interplay between the two DWs when they are close enough. Subsequently, the two DWs connect and shift with a same speed, as clearly shown in the supplementary material Movie.S1 which gives the motion of the DWs for $B_N = 0.03$. In some extent, this phenomenon can be understood qualitatively by the analogue between the DWs interplay and an inelastic collision of two quasiparticles, as detailed explained in earlier work.[18] Interestingly, the AFM2 DW detaches from the AFM1 DW when $B_N$ further increases above the critical value ~ 0.048, as clearly shown in the supplementary material Movie.S2 and Movie.S3 which present the local magnetization and spin configuration for $B_N = 0.05$, respectively. It is clearly shown that the AFM1/AFM2 DW rotates by 180 degree around the $x$-/$z$-axis after the detachment.

The equilibrium velocities of the DWs for various $B_N$ are summarized in Fig. 3(a). On the one hand, the velocity increases nonlinearly with the increase of $B_N$ (below the critical value) due to the spin-wave excitation as shown in Fig. 3(b) which gives the local magnetization for $B_N = 0.04$. It is noted that the spin-wave excitation acting as an additional energy dissipation is enhanced as $B_N$ increases, resulting in the decrease of the acceleration with $B_N$, as confirmed in our simulations. As a matter of fact, the AFM2 DW is driven by the AFM1 DW through the damping torque ~ $-\mathbf{S} \times (\mathbf{S} \times \mathbf{H}_{inter})$ resulted from the coupling between the two DWs, as depicted in Fig. 3(c). When $B_N$ is increased, the two DWs get close to each other, resulting in the increase of the damping torque. The maximum value of the torque is obtained at the critical $B_N$ with the positions of the two DWs coincide with each other (middle layer in Fig. 3(c)), and the AFM2 DW also reaches its highest possible velocity $v_c$. For this case, the coupling energy between the two DWs is rather high, reducing the stability of the DWs. As $B_N$ further increases, the wall linking is not stable and the detachment occurs. Furthermore, the spins in the AFM1/AFM2 DW are driven out of the easy plane and rotate by 180 degree around the $x$-/$z$-axis to save the interface exchange coupling energy.

Subsequently, we investigate the effects of the interaction couplings on the spin dynamics and the critical velocity $v_c$. Fig. 4(a) gives the velocities of the DWs (solid and empty circles) under $B_N = 0.04$ and $v_c$ (blue asterisks under labeled critical $B_N$) for various $J_{AF2}$ at $J_{inter} = 0.5$.

It is well noted that the energy of the DW determines its stabilization, i. e., higher energy results in more stable DW. Thus, the AFM2 DW is further stabilized as $J_{AF2}$ increases, leading to the increase of the critical $B_N$ and $v_c$, as clearly shown in our simulations. Furthermore, for $J_{AF2} > 1$, the velocities of the DWs under $B_N = 0.04$ slightly decrease with the increasing $J_{AF2}$ due to the reduction of their mobility. Furthermore, the effects of $J_{inter}$ are also studied, and the corresponding results are presented in Fig. 4(b). Both $v_c$ and the critical $B_N$ sharply increase as $J_{inter}$ increases, and then slowly increase for $J_{inter} > 0.5$. Thus, it is suggested that $v_c$ is mainly determined by $J_{AF2}$ for a rather large $J_{inter}$.

As a matter of fact, biaxial anisotropy does exist in real materials and is believed to play an essential role in the current-induced orientation of the AFM order.[31] Thus, we also studied the effects of the intermediate anisotropies $d_z$ and $D_x$ on the AFM dynamics for integrity. Fig. 4(c) gives the simulated velocity as a function of $d_z$ for various $D_x$ under $B_N = 0.04$. It is clearly shown that the velocity is increased with the increasing $d_z$ and/or $D_x$, demonstrating the important role of the intermediate anisotropy in the motion of AFM DWs. One may note that the energy gap between the AFM DW and domain prominently determines the mobility of the DW. When the intermediate anisotropy is considered, the energy gap is reduced which enhances the reversal of the local spins in the domain and results in higher mobility of the DW. Thus, both $d_z$ and $D_x$ can speed up the AFM DWs, as confirmed in our simulations. Furthermore, both $v_c$ (asterisks) and the critical $B_N$ increase due to the increases of the DW energy when $d_z$ and/or $D_x$ is increased. For example, the critical velocity for $d_z = 0.01$ and $D_x = 0.05$ (red asterisk) increase by ~25% comparing with the uniaxial case.

At last, the effects of the damping constants on the velocities are also investigated and the simulated results are presented in Fig. 4(d) which gives the velocities as functions of $\alpha_2$ for various $\alpha_1$ under $B_N = 0.04$. It is clearly shown that the velocity decreases as the damping constant increases. One may note that the damping torque $\boldsymbol{\Gamma}_{ex}^{\alpha}$ (Fig. 1(a)) is enhanced with the increase of the damping constant, which reduces the precession torque $\boldsymbol{\Gamma}_{ex}^{\eta}$. As a result, the effects of $B_N$ are significantly suppressed, which speeds down the AFM DWs. Furthermore, the critical $B_N$ is increased, while $v_c$ significantly decreases with the increase of the damping constant, as shown in our simulations.

So far, it has been proven that the AFM2 DW can be efficiently driven by the AFM1 DW

through the interface coupling, and the two DWs will be detached when $B_N$ increases above a critical value. As a matter of fact, these phenomena could be used in future AFM spintronics device design. For example, the bidirectional control of AFM2 multi DWs can be realized through elaborately modulating $B_N$ on the AFM1 layers. The proposed structure is shown in Fig. 5(a) in which the different domains (blue and red arrays in the top layer) are used to store information bits (0 or 1) and their lengths determine the bit numbers. Furthermore, the other AFM layer (bottom layer) with single DW under $B_N$ is used to be a driving bar. The multi DWs can be driven by alternately applying small and large $B_N$, as depicted in the insert of Fig. 5. Specifically, the first AFM2 DW can be driven by the driving DW under a small $B_N$, and detaches with the driving DW (middle layer in the inset of Fig. 5) when $B_N$ increases above the critical value. Subsequently, a small $B_N$ is used to drive the motion of the second AFM2 DW. As a result, the multi DWs can be indirectly driven by the driving AFM DW, as shown in Fig. 5(b). Then, The driving DW can shift back to the initial position by applying a large opposite $B_N$ as depicted in Fig. 5(c). Moreover, the reverse motion of the multi DWs can be driven by the opposite $B_N$ through the inverse processes, and the picture is not shown here for brevity.

Up to now, there is an urgent need in modulating AFM multi DWs to provide useful information for practical applications. In this work, the AFM DW has been proven to be driven efficiently by the other AFM DW under the staggered field through the interface coupling in heterostructures. Furthermore, the highest possible velocity of the AFM DW relevant to several factors has been clarified and detailed explained. Finally, the control of multi DWs is proposed based on the detachment of the two DWs under the staggered field larger than the critical value.

**Acknowledgements**:

The work is supported by the National Key Projects for Basic Research of China (Grant No. 2015CB921202), and the Natural Science Foundation of China (Grant No. 51332007), and the Science and Technology Planning Project of Guangdong Province (Grant No. 2015B090927006), and the Natural Science Foundation of Guangdong Province (Grant No. 2016A030308019).

**FIGURE CAPTIONS**

Figure 1. (color online) (a) Illustration of torques exerted on the center of AFM DW, and (b) schematic illustration of AFM1 and AFM2 DWs in the heterojunction.

Figure 2. (color online) (a) The initial spin configuration, and (b) the transverse component of AFM1 and AFM2 DWs. (c) The displacement of the DWs dependent of time under various staggered fields. The position of AFM1/AFM2 DW is marked by the solid/dashed line.

Figure 3. (color online) (a) the velocities of the DWs as functions of the staggered field $B_N$, and (b) the components of the local magnetization near the attached DWs under $B_N = 0.04$, and the calculated $n_1^z$ and $n_2^x$ demonstrate an spin-wave emission, and (c) the depictions of the spin configurations of the DWs under various staggered fields.

Figure 4. (color online) The velocities of the DWs (solid and dashed circles) under $B_N = 0.04$ as functions of (a) $J_{AF2}/J$, and (b) $J_{inter}/J$. The critical $B_N$ and velocity are also presented by the blue number and asterisks, respectively. The velocities of the DWs as functions of (c) $d_z$ for various $D_x$ for $\alpha_1 = \alpha_2 = 0.01$, and (d) $\alpha_2$ for various $\alpha_1$ for $d_z = D_x = 0.02$. The critical $B_N$ and velocity (asterisks) for several parameters are also given, as an example.

Figure 5. (color online) The proposed driving mechanisms for multi DWs motion. (a) The different domains (blue and red arrays in the top layer) are used to store information bits (0 or 1) and their lengths determine the bit numbers. The AFM layer (bottom layer) with single DW under $B_N$ is used to be a drive bar. (b) The multi DWs motion driven by an alternating small and large $B_N$, and (c) the DW in drive bar can shift back to the initial position by applying a large opposite $B_N$.

**SUPPLEMENTARY MATERIALS CAPTIONS**

Movie.S1. The motion of the AFM DWs for $B_N = 0.03$. The used simulation parameters are $J_{AF1} = J_{AF2} = J$, $J_{inter} = 0.5J$, $d_x = D_z = 0.1J$, and $\alpha_1 = \alpha_2 = 0.01$.

Movie.S2. The motion of the AFM DWs for $B_N = 0.05$. The used simulation parameters are $J_{AF1} = J_{AF2} = J$, $J_{inter} = 0.5J$, $d_x = D_z = 0.1J$, and $\alpha_1 = \alpha_2 = 0.01$.

Movie.S3. The spin configuration in the motion of AFM DWs for $B_N = 0.05$. The used simulation parameters are $J_{AF1} = J_{AF2} = J$, $J_{inter} = 0.5J$, $d_x = D_z = 0.1J$, and $\alpha_1 = \alpha_2 = 0.01$.

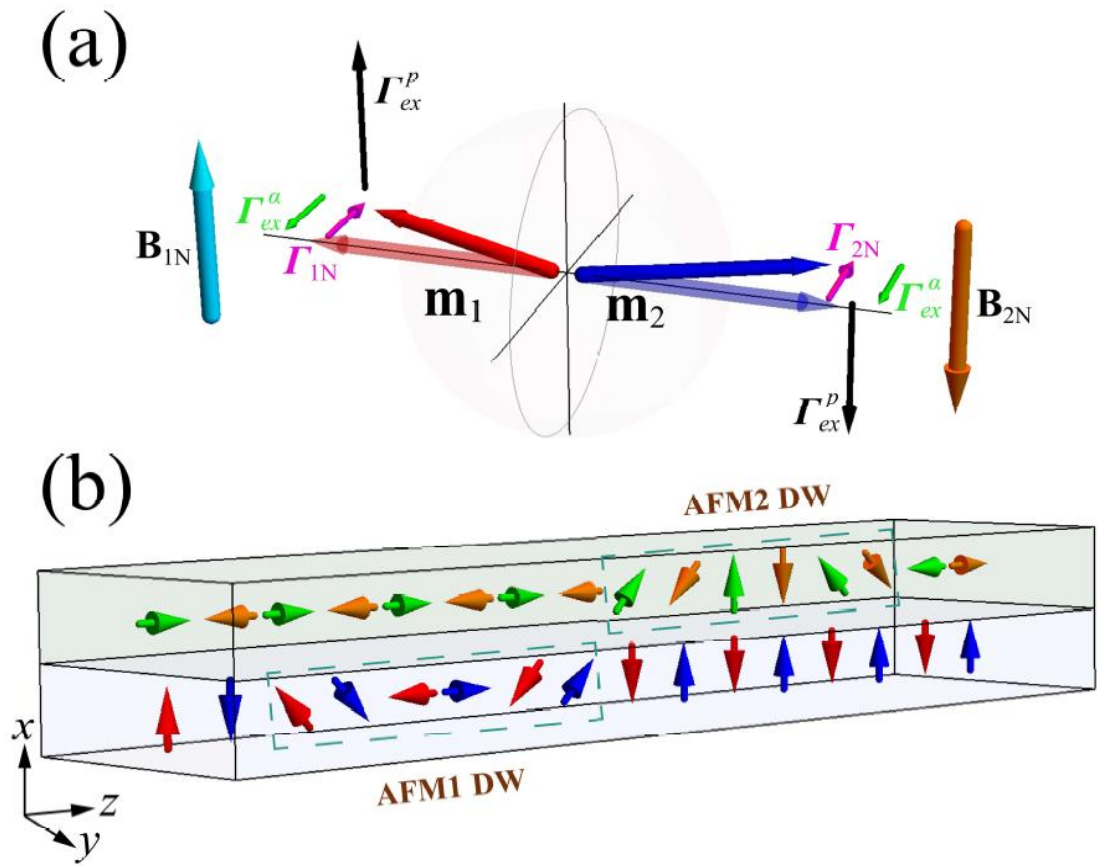

Figure 1. (color online) (a) Illustration of torques exerted on the center of AFM DW, and (b) schematic illustration of AFM1 and AFM2 DWs in the heterojunction.

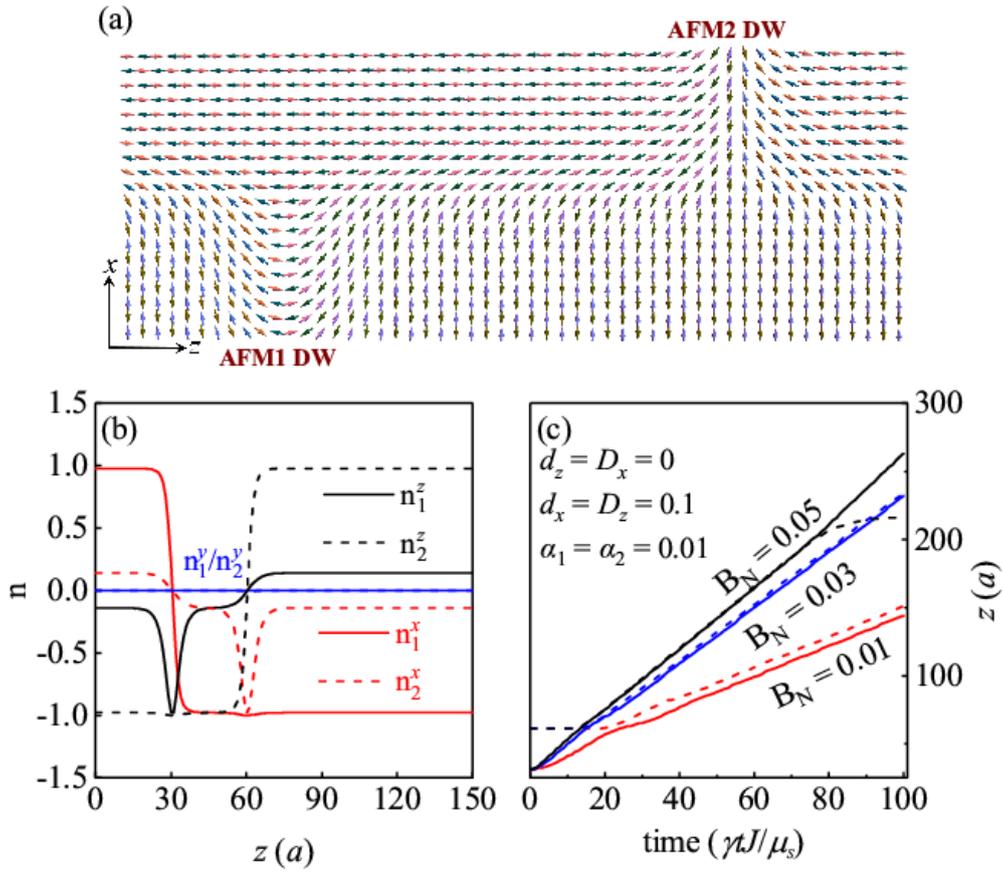

Figure 2. (color online) (a) The initial spin configuration, and (b) the transverse component of AFM1 and AFM2 DWs. (c) The displacement of the DWs dependent of time under various staggered fields. The position of AFM1/AFM2 DW is marked by the solid/dashed line.

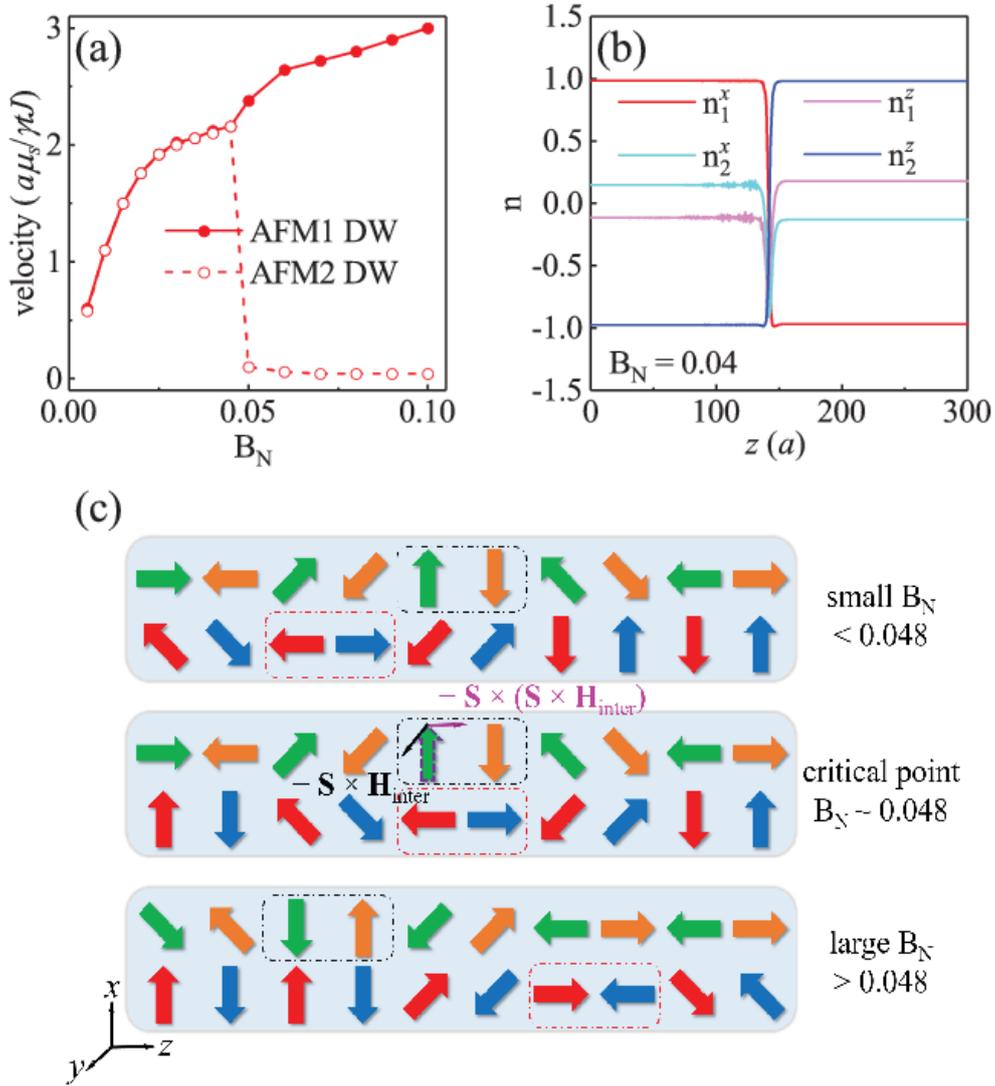

Figure 3. (color online) (a) the velocities of the DWs as functions of the staggered field $B_N$, and (b) the components of the local magnetization near the attached DWs under $B_N = 0.04$, and the calculated $n_1^z$ and $n_2^x$ demonstrate an spin-wave emission, and (c) the depictions of the spin configurations of the DWs under various staggered fields.

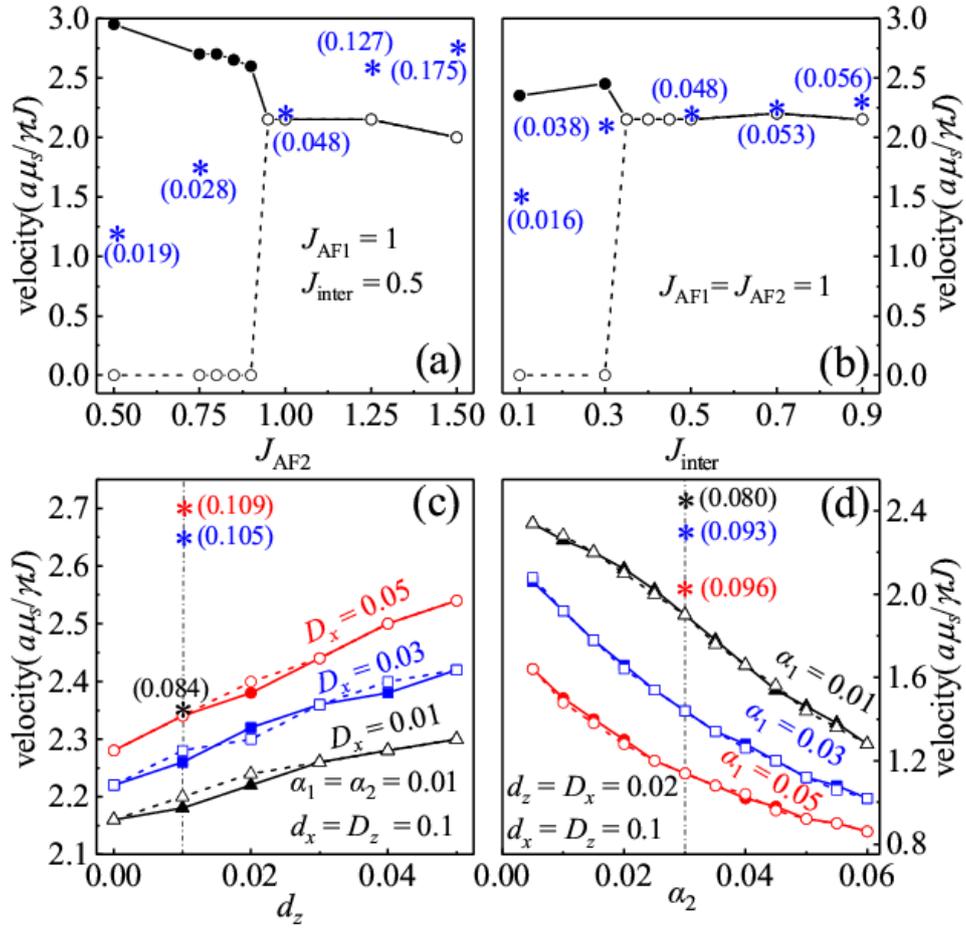

Figure 4. (color online) The velocities of the DWs (solid and dashed circles) under $B_N = 0.04$ as functions of (a) $J_{AF2}/J$, and (b) $J_{inter}/J$. The critical $B_N$ and velocity are also presented by the blue number and asterisks, respectively. The velocities of the DWs as functions of (c) $d_z$ for various $D_x$ for $\alpha_1 = \alpha_2 = 0.01$, and (d) $\alpha_2$ for various $\alpha_1$ for $d_z = D_x = 0.02$. The critical $B_N$ and velocity (asterisks) for several parameters are also given, as an example.

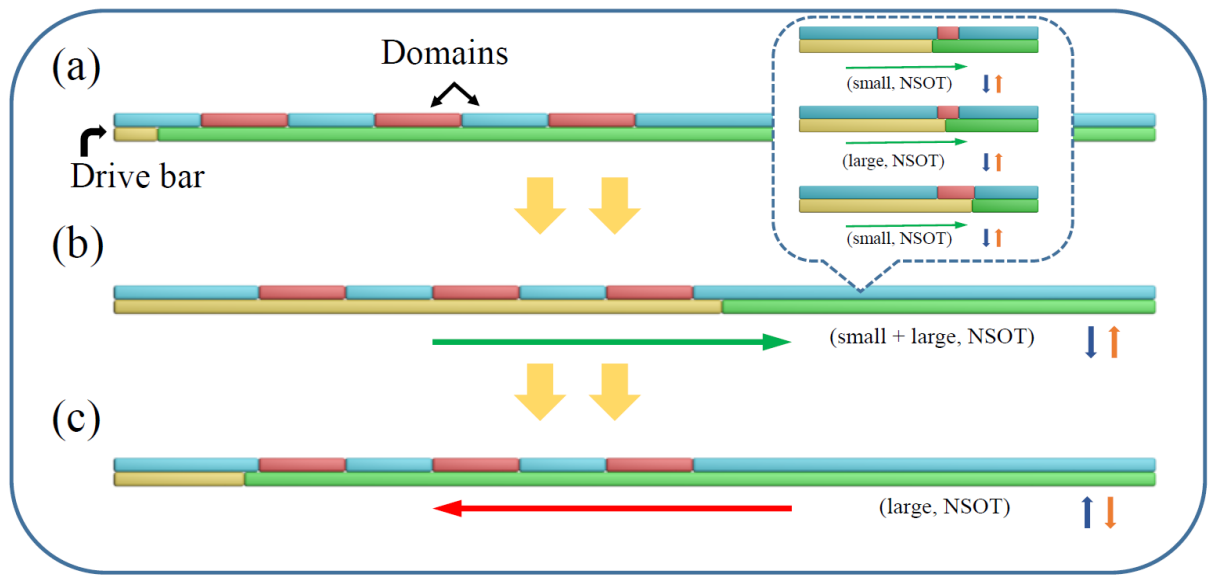

Figure 5. (color online) The proposed driving mechanisms for multi DWs motion. (a) The different domains (blue and red arrays in the top layer) are used to store information bits (0 or 1) and their lengths determine the bit numbers. The AFM layer (bottom layer) with single DW under $B_N$ is used to be a drive bar. (b) The multi DWs motion driven by an alternating small and large $B_N$, and (c) the DW in drive bar can shift back to the initial position by applying a large opposite $B_N$.